\documentclass{PoS}

\title{The deconfinement phase transition in the Hamiltonian approach to Yang--Mills theory in Coulomb gauge}

\ShortTitle{Deconfinement phase transition in the Hamiltonian approach to YM-theory}

\author{\speaker{Hugo Reinhardt}\\
Universit\"at T\"ubingen, Institut f\"ur Theoretische Physik\\
Auf der Morgenstelle 14, 72076 T\"ubingen, Germany\\
E-mail: \email{hugo.reinhardt@uni-tuebingen.de}}

\author{Davide R.~Campagnari and Jan Heffner\\
Universit\"at T\"ubingen, Institut f\"ur Theoretische Physik\\
Auf der Morgenstelle 14, 72076 T\"ubingen, Germany
}

\abstract{
The deconfinement phase transition of SU$(2)$ Yang--Mills theory is investigated in the Hamiltonian approach in Coulomb gauge assuming
a quasi-particle picture for the grand canonical gluon ensemble. The thermal equilibrium state is found by minimizing the free energy with
respect to the quasi-gluon energy. At the deconfinement phase transition the gluon energy, being infrared divergent in the confined phase,
becomes infrared finite in the deconfined phase, while the ghost form factor remains infrared divergent in the deconfined phase but its infrared
exponent is approximately halved. Using the lattice results for the gluon propagator to fix the scale the deconfinement
transition temperature is obtained in the range of $275$ to $290$ MeV.}

\FullConference{Xth Quark Confinement and the Hadron Spectrum,\\
		October 8-12, 2012\\
		TUM Campus Garching, Munich, Germany}

\usepackage{amsmath}

\newcommand{\ii}{\mathrm{i}}

\newcommand{\be}{\begin{equation}}
\newcommand{\ee}{\end{equation}}
\newcommand{\bea}{\begin{eqnarray}}
\newcommand{\eea}{\end{eqnarray}}

\newcommand{\rk}{\right)}
\newcommand{\lk}{\left(}

\DeclareMathOperator{\Det}{Det}
\DeclareMathOperator{\Tr}{Tr}

\renewcommand*{\vec}[1]{#1}
\newcommand{\vB}{\vec{B}}
\newcommand{\vA}{\vec{A}}
\newcommand{\vx}{\vec{x}}
\newcommand{\vy}{\vec{y}}
\newcommand{\vp}{\vec{p}}
\providecommand*{\coloneq}{\mathrel{\mathop:}=}
\newcommand*{\dd}[1][]{\mathop{\mathrm{d}^{#1}}\mkern-4mu}

\newcommand{\ket}[1]{\lvert#1\rangle}
\newcommand{\braket}[2]{\langle #1 \vert #2 \rangle}
\newcommand*{\vev}[1]{\left< #1 \right>}

\begin{document}
\section{Introduction}

Understanding the deconfinement phase diagram of QCD is one of the major challenges of particle physics. In this talk I will report on a 
description of the deconfinement phase transition at zero baryon density within the Hamiltonian approach to Yang--Mills theory in Coulomb gauge, 
extending the previously developed variational approach for the Yang--Mills vacuum \cite{FeuRei04} to finite temperatures. 
This approach has been quite successful in understanding the infrared properties of Yang--Mills theory: One finds a gluon energy which is infrared divergent  \cite{FeuRei04} and a linearly rising quark potential \cite{EppReiSch07}, both being signals of confinement. In addition, one finds an infrared enhanced running coupling constant with no Landau pole \cite{SchLedRei06}, a perimeter law for the 't~Hooft loop \cite{ReiEpp07} and, within an approximate Dyson--Schwinger equation, an area law for the spatial Wilson loop \cite{Pak:2009em}. Also the topological susceptibility was found in accord with the lattice data \cite{Campagnari:2008yg}. Given the success of this approach in the vacuum sector we can also expect a decent description of the deconfinement phase transition. The outline of my talk is 
as follows: I will first review the basic ingredients of the Hamiltonian approach to Yang--Mills theory in Coulomb gauge. Within this approach I will then study
the grand canonical ensemble in a mean-field type approximation and investigate the deconfinement phase transition. 

\section{Hamilton approach to Yang--Mills theory}

The Hamilton approach to Yang--Mills theory is based on the canonical quantization in Weyl gauge $A_0 = 0$, which yields the Hamiltonian
\be
\label{1}
H = \frac{1}{2} \int \dd^3 x \lk \vec{\Pi}^2 (\vx) + \vec{B}^2 (\vx) \rk \, ,
\ee
where $\vec{\Pi} (x) = - \ii \delta / \delta \vA (\vx)$ is the canonical momentum operator. Due to the use of the Weyl gauge Gauss' law 
escapes the quantum equations of motion and has to be imposed as a constraint to the wave functional
\be
\label{138}
D \Pi \psi [A] = 0 \, .
\ee
Here $D$ denotes the covariant
derivative in the adjoint representation of the gauge group. The operator in Gauss' law $D \Pi$ is nothing but the generator of (time-independent
but space-dependent) gauge transformations and Gauss' law ensures that (in the absence of matter fields) the wave functional must be 
gauge invariant. To respect gauge invariance one can work with explicitly gauge invariant wave functionals, which has been pursued mainly
in $2+1$-dimensions but which becomes exceedingly involved in $3+1$-dimensions. A more convenient way is to fix the gauge and resolve 
Gauss'~law explicitly. For this purpose Coulomb gauge is convenient and after gauge fixing one finds the following Hamiltonian
\be
\label{3}
H = \frac{1}{2} \int \dd^3 x \lk J_A^{- 1} \vec{\Pi} (\vx) J_A \cdot \vec{\Pi} (\vx) + \vB^2 (\vx) \rk + H_\text{C} \, ,
\ee
where $J_A = \Det (- D \partial)$ is the Faddeev--Popov determinant and 
\be
\label{4}
H_\text{C} = \frac{g^2}{2} \int \dd^3 x \, \dd^3 y\,J_A^{- 1} \, \rho^a (\vx) J_A \, F^{ab} (\vx, \vy) \rho^b (\vy)
\ee
is the so-called Coulomb Hamiltonian. Here, $\rho^a = - f^{abc} A^b \Pi^c$ is the color charge density of the gluons. The gauge fixed Hamiltonian
is highly non-local due to the presence of the Faddeev--Popov determinant and the Coulomb term $H_\text{C}$. Although the gauge fixing gives rise to 
a more complicated (non-local) Hamiltonian it has the advantage that the gauge invariance has been taken care of once and for all and that after
Coulomb gauge fixing any wave functional depending on the transversal gauge field only lies in the physical Hilbert space. Fortunately it 
turns out that in the gluon sector, in particular for the investigation of the infrared properties, the Coulomb term can be ignored, which I will
do in the following.

With the Coulomb gauge fixed Hamiltonian (\ref{3}) the Yang--Mills Schr\"odinger equation has been solved by a variational principle using 
Gaussian type ans\"atze for the vacuum wave functional. The approach developed in our group differs from previous attempts \cite{Schutte:1985sd, Szczepaniak:2001rg}  by the ansatz
for the vacuum wave functional, by the full inclusion of the Faddeev--Popov determinant and by the renormalization. Our ansatz for the vacuum
wave functional reads 
\be
\label{5}
\psi [A] = J_A^{- 1/2} \exp \left[ - \frac{1}{2} \int A \omega A \right] \, ,
\ee
where $\omega$ is a variational kernel, which is determined by minimizing the energy $\langle \psi | H | \psi \rangle$. The pre-exponential factor
has the advantage that it cancels the Faddeev--Popov determinant in the scalar product of the Coulomb gauge fixed wave functionals. Furthermore,
for this wave functional the gluon propagator is given by 
\begin{figure}[t]
\centering
 \includegraphics[width=0.45\linewidth]{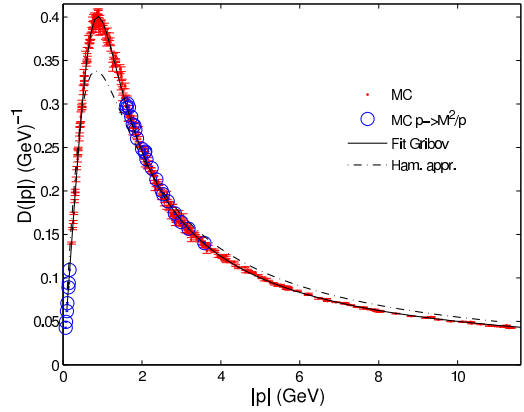}
\caption{Comparison of the results of the variational calculation and lattice data for the gluon propagator.}
\label{fig:lattice}
\end{figure}
\be
\label{6}
\langle A A \rangle = (2 \omega)^{- 1} \,,
\ee
which identifies the Fourier transform of $\omega$ as quasi-gluon energy. Figure \ref{fig:lattice} shows the result of the variational calculation
for the gluon propagator in comparison with the lattice data. As one observes, the cross feature of the lattice data are well reproduced. In 
particular, in the deep infrared regime our variational calculations agree perfectly with the lattice data. There are deviations in the mid-momentum
regime, which can be substantially reduced by using non-Gaussian wave functionals, Ref.~\cite{CamRei10}. What is also remarkable is that the lattice
data can be nicely fitted by Gribov's formula 
\be
\label{185-GY}
\omega (p) = \sqrt{\vp^2 + M^4 / \vp^2} 
\ee
with the mass $M = 0.88$ GeV.

\section{The grand canonical ensemble}

Since the gluons have vanishing chemical potential their grand canonical ensemble is defined by the density operator
\be
\label{8}
D = \exp \lk - \beta H \rk \, ,
\ee
where $H$ is the gauge fixed Hamiltonian (\ref{3}) and $\beta = 1/k_B T$ is the inverse temperature. By means of the density matrix thermal 
averages are defined by
\be
\label{9}
\langle \ldots \rangle_T \coloneq \frac{\Tr (D \ldots)}{\Tr D} \, ,
\ee
where the trace can be calculated, in principle, in any complete basis. However, we will necessarily have to introduce 
approximations and then the choice of the basis matters. We will choose an optimal basis by exploiting the variational principle. Inspired by the zero temperature calculations, see Eq.~(\ref{5}), we will choose our basis
in the form
\be
\label{10}
\ket {\tilde{k} } = \frac{1}{\sqrt{\Det (- D \partial)}} \ket{ k } \, ,
\ee
where the states $| k \rangle$ are defined by the Fock basis 
\be
\label{225-GX}
\ket{ 0} \, , \quad a^\dagger (\vp) \ket{ 0}  \, , \quad a^\dagger (\vp_1) a^\dagger (\vp_2) \ket{ 0} ,\quad \ldots 
\ee
obtained by decomposing the gauge field in terms of creation-annihilation operators
\be
\label{11}
A (k) = \frac{1}{\sqrt{2 \omega (k)}} \lk a (k) + a^\dagger (-k) \rk \, 
\ee
with a so far arbitrary kernel $\omega (k)$ and the vacuum state, being defined by $a (k) | 0 \rangle = 0$, is the Gaussian 
\be
\label{13}
\braket {A }{ 0 } = \exp \lk - \frac{1}{2} \int A \omega A \rk \, .
\ee
Inserting this state into Eq.~(\ref{10}) we find that the state $| \tilde{0} \rangle$ is nothing but the trial ansatz (\ref{5}) 
for the zero temperature
variational calculation. However, in the present case the kernel $\omega (k)$ is not determined by minimizing the energy but at the moment an 
arbitrary kernel. 

We cannot treat the full density operator, Eq.~(\ref{8}). Therefore we replace in the density operator the full Hamiltonian (\ref{4}) 
by a single-particle one
\be
\label{14}
D = \exp (- \beta h) \, , \quad h = \int \dd p\, \epsilon (p) a^\dagger(p) a (p) \, .
\ee
As a consequence, for the thermal averages (\ref{9}) Wick's theorem applies, which tremendously simplifies the calculations of thermal expectation
values. With the density matrix (\ref{14}) one finds the usual Bose occupation numbers
\be
\label{15}
\vev { a^\dagger (p) a (p) } \equiv n (p) = \left( \exp (\beta \epsilon (p)) - 1 \right)^{- 1} \, ,
\ee
while the finite-temperature gluon propagator is given by
\be
\label{16}
\langle A A \rangle_T = (1 + 2n) / (2 \omega) \, .
\ee
It differs from the zero-temperature propagator (\ref{6}) only by the presence of the finite-temperature occupation numbers $n (p)$. 
In the limit
of a vanishing temperature $T \to 0$ the $n (p)$ vanish and Eq.~(\ref{16}) reduces to the zero-temperature propagator,
Eq.~(\ref{6}).

With the density operator (\ref{14}) at hand, we can straightforwardly calculate the entropy 
\be
\label{17}
S = - k_B \Tr D \ln D
\ee
and the free energy
\be
\label{18}
F [\epsilon, \omega] = \langle H \rangle_T - TS \, .
\ee
A comment is here in order: We cannot calculate in here the free energy from the partition function $Z = \Tr D$ since the density matrix $D$ (\ref{14})
is not yet known. Rather we calculate the free energy by taking the thermal expectation value of the full Hamiltonian. This has in addition
the advantage that also two-body correlations are included, which cannot be captured by the single-particle density matrix (\ref{14}).

\section{The deconfinement phase transition}

\begin{figure}[t]
\centering
\parbox[t]{.45\linewidth}{
\centering
 \includegraphics[width=0.5\linewidth]{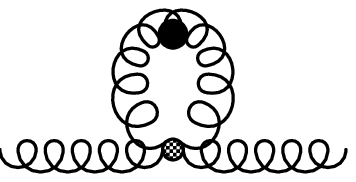}
\caption{Tadpole term $I[n]$.}
\label{fig:tadpole}
}
\parbox[t]{.45\linewidth}{
 \includegraphics[width=\linewidth]{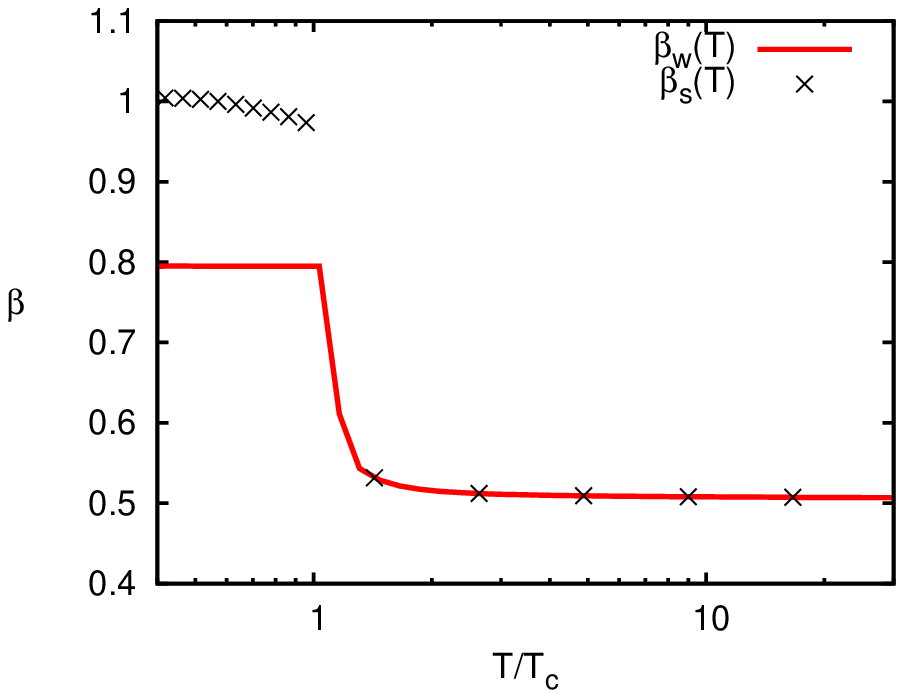}
\caption{Infrared exponent $\beta$ of the ghost form factor as a function of temperature.}
\label{fig:beta}
}
\end{figure}

So far, we have two unknown kernels. The single-particle energies $\epsilon (p)$ occurring in the single-particle density operator (\ref{14})
and the kernel $\omega (p)$ occurring in the vacuum state (\ref{13}) of our Fock basis (\ref{225-GX}), (\ref{13}). We now determine these kernels by 
minimizing the free energy (\ref{18}). Since $n (p)$ is a monotonous function of $\epsilon (p)$ we can vary $F [\epsilon, n]$ with respect
to $n (p)$, which yields the relation \cite{HefReiCam12}
\be
\label{19}
\epsilon  (p) = \omega (p) \, .
\ee
This relation is not surprising given the form of the thermal gluon propagator (\ref{16}), which identifies $\omega (p)$ as quasi-gluon energy. Let us 
stress, however, that if one goes beyond the present approximation and includes also the Coulomb term (\ref{4}), one obtains still a linear relation between
$\epsilon (p)$ and $\omega (p)$, however, the proportionality factor differs from one by an additional loop integral \cite{Reinhardt:2011hq}. Finally, variation with respect
to the choice of our basis, i.e. with respect to $\omega (p)$ yields the gap equation
\be
\label{20}
\omega^2 (p) = p^2 + \chi^2 (p) + I [n] \,,
\ee
where $\chi (p)$ denotes the ghost loop calculated from the finite-temperature gluon propagator and $I [n]$ denotes the tadpole of the transversal spatial gluons, see Fig.~\ref{fig:tadpole}.
The gap equation (\ref{20}) has to be solved together with the Dyson--Schwinger equation for the ghost form factor $d (p)$. These equations can be
solved analytically in the infrared by power law ans\"atze 
\be
\label{22}
\omega (p) = \frac{A}{p^\alpha} \, , \quad d (p) = \frac{B}{p^\beta} \, .
\ee
Assuming the horizon condition $d^{- 1} (p = 0) = 0$ one finds at zero temperature from the ghost DSE the sum rule
\be
\label{23}
\alpha = 2 \beta + 2 - d \, ,
\ee
where $d$ is the number of spatial dimensions. With this sum rule one finds then from the gap equation the following solutions for the infrared 
exponent $\beta$
\begin{align}
\label{24}
d & = 3\,:\quad \beta = 1.00 \,, \quad \beta \approx 0.795\nonumber\\
d & = 2\,:\quad \beta = 0.50 \, .
\end{align}
The same exponents are extracted numerically from the self-consistent solution of the equations of motion.

\begin{figure}[t]
\includegraphics[width=.45\linewidth]{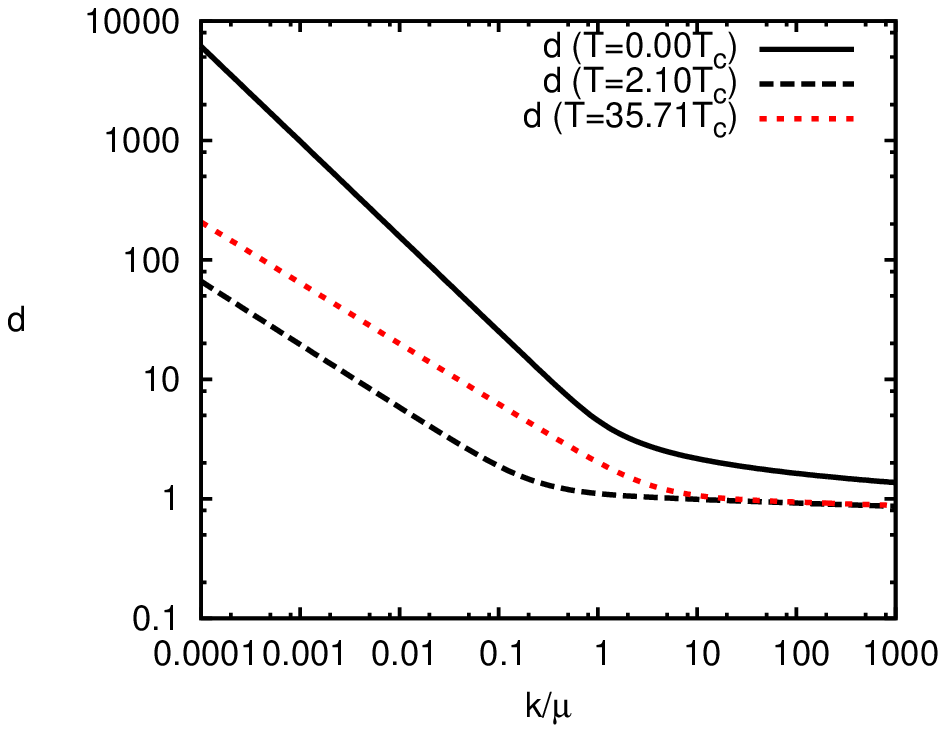}\hfil
\includegraphics[width=.45\linewidth]{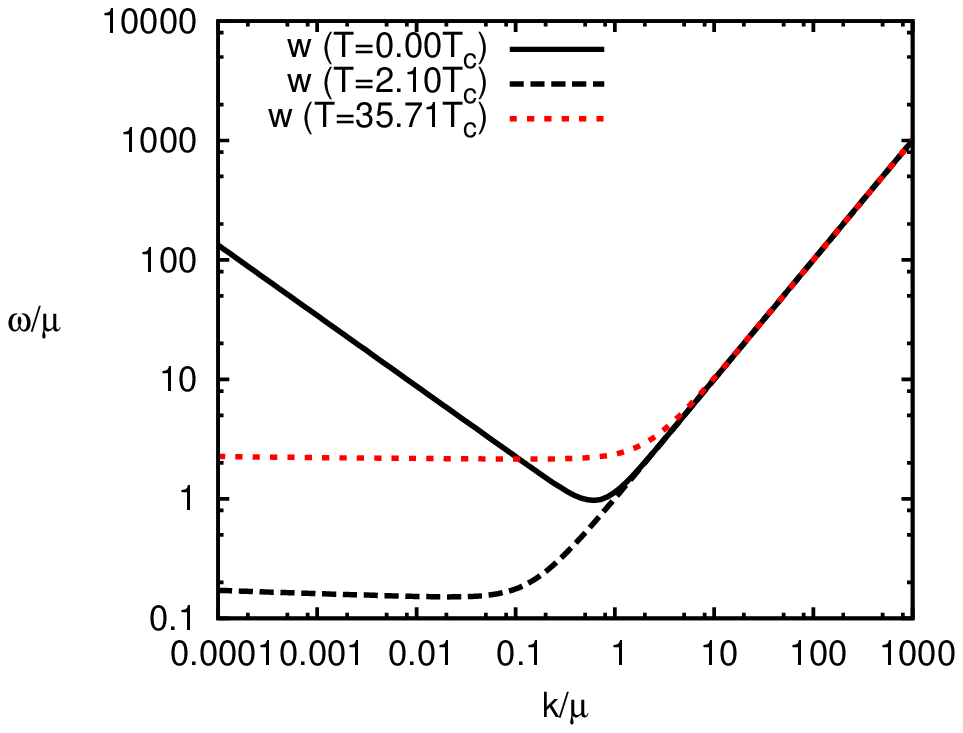}
\caption{Zero- and finite-temperature solutions for the ghost form factor $d(p)$ (left panel) and the gluon kernel $\omega(p)$ (right panel).}
\label{fig:fTsol}
\end{figure}

\begin{figure}[t]
\centering
\parbox[t]{.45\linewidth}{
 \includegraphics[width=\linewidth]{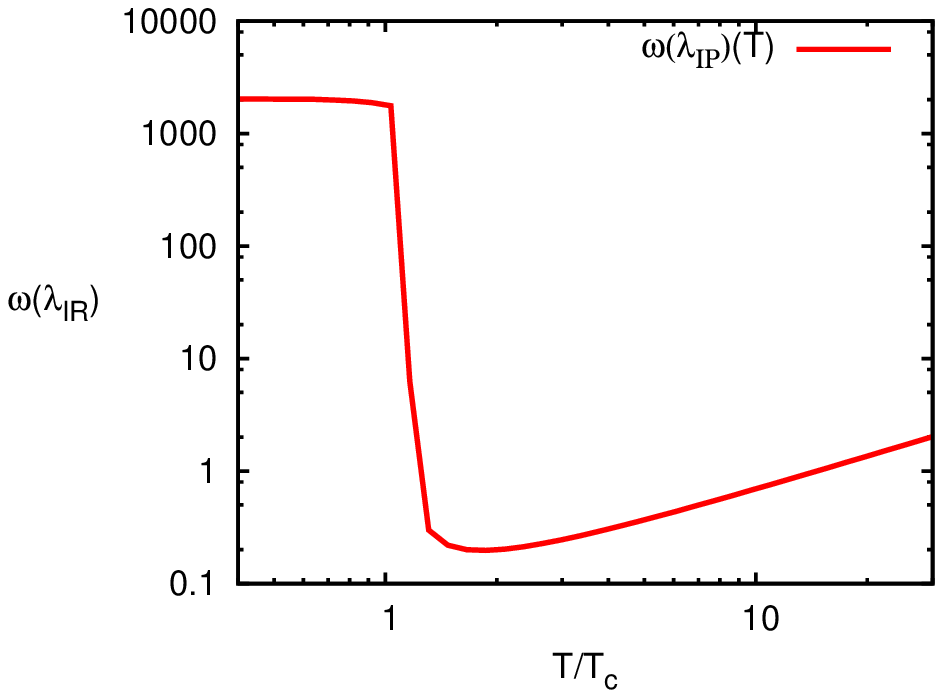}
\caption{The effective gluon mass $\omega (\lambda_{IR})$ as a function of temperature.}
\label{fig:effmass}
}
\hfil
\parbox[t]{.45\linewidth}{
 \includegraphics[width=\linewidth]{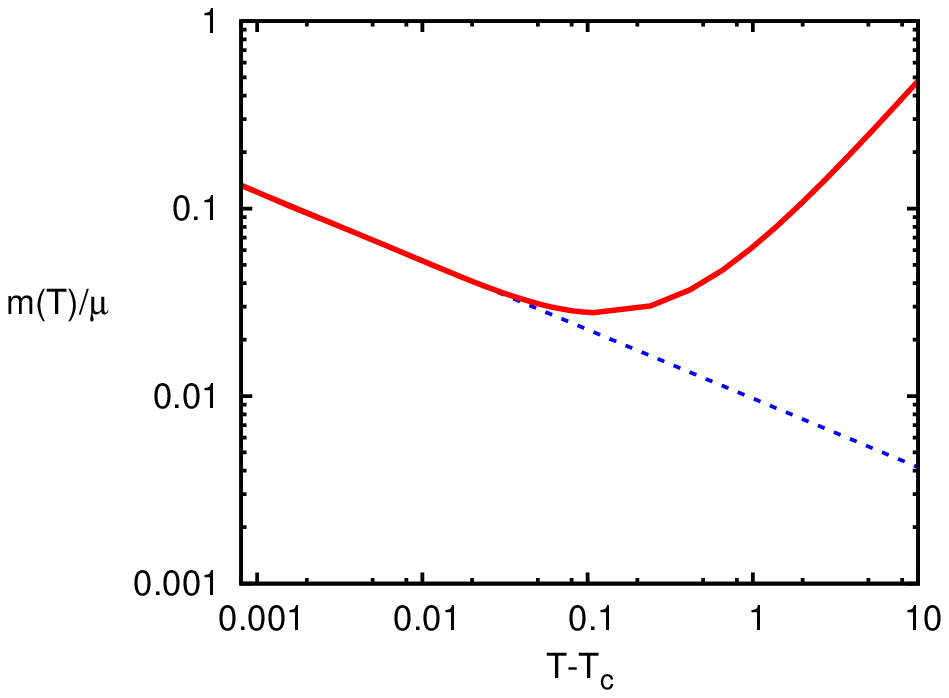}
\caption{Critical behavior of the effective gluon mass for $T \to T_c$.}
\label{fig:critmass}
}
\end{figure}

At arbitrarily finite temperature an infrared analysis is impossible due to the fact that the gluon energy $\omega (p)$ occurs in the occupation
number (\ref{15}) in exponential form. One can, however, carry out the infrared analysis at infinite temperature, where the Bose occupation numbers given
by Eqs.~(\ref{15}), (\ref{19}) reduce to 
\be
\label{25}
T \to \infty\,:\quad n (p) \to (\beta \omega (p))^{-1} \, .
\ee
One finds then the same sum rule (\ref{23}) as at zero temperature and a single solution with
\be
\label{26}
T \to \infty \,: \quad d = 3 \, : \quad\beta = 0.5 \, , \quad \alpha = 0 \, .
\ee
Figure \ref{fig:beta} shows the infrared exponent of the ghost form factor $\beta$ as function of temperature. At low temperature one finds the two 
solutions for the infrared exponents obtained from the zero-temperature infrared analysis. These exponents stay more or less constant as the
temperature increases up to a critical temperature, where the two solutions for the infrared exponent $\beta$ merge to a single one, which with
increasing temperature approaches the value $\beta = \frac{1}{2}$ predicted by the infrared analysis in the high temperature limit. Figure~\ref{fig:fTsol}
shows the numerical solution for the ghost form factor $d (p)$ and the gluon energy $\omega (p)$ for various temperatures. The obtained results
are in agreement with the predictions of the infrared analysis. At zero temperature both $d (p)$ and $\omega (p)$ are infrared diverging. Above
$T_c$ the ghost form factor is still infrared diverging but its infrared exponent is halved while the gluon energy becomes infrared finite in 
accordance with the prediction of the infrared analysis (\ref{26}). Furthermore above $T_c$ the plateau value of the gluon energy
in the infrared increases with the temperature. This is seen in Fig.~\ref{fig:effmass}, where we show the gluon energy at an infrared cut-off $\lambda_{IR}$. 
$\omega (\lambda_{IR}) = m (T)$ jumps at the deconfinement phase transition from a very large to a small value and afterwards increases 
linearly with the temperature. Zooming into the transition regime, which is done in Fig.~\ref{fig:critmass} one can extract the critical exponent of 
$\omega (\lambda_{IR}) = m (T)$  defined by 
\be
\label{27}
m_{crit} (T) \sim \lk T/{T_C} - 1 \rk^\kappa  \, 
\ee
and obtains the value $\kappa  = 0.37$. This value compares well with the value $\kappa = 0.41$ obtained in Ref.~\cite{Castorina:2011ja} in a phenomenological quasi-particle
model for the gluons using input data from the $d = 3$ Ising model, which is in the same universality class as SU$(2)$ gauge theory. Fitting our
scale at the value of the Gribov mass $M = 0.88$ GeV, see Eq.~(\ref{185-GY}), determined in Ref.~\cite{BurQuaRei09} on the lattice, we find a critical temperature
for the deconfinement phase transition of $T_c = 275 \ldots 290$ MeV. This is in reasonable agreement with the lattice value of $T_c = 295$ MeV.
This result is also in reasonable agreement with the value $T_c \simeq 270$ MeV obtained from the effective potential of the Polyakov loop
calculated in Ref.~\cite{ReiHef12} within the Hamiltonian approach. Let me also mention that recently the Hamilton approach to Yang--Mills theory in 
Coulomb gauge has been also used to study glueballs at finite temperature \cite{YepezMartinez:2012rf}.
The results obtained in the gluon sector at finite temperature are quite encouraging for an extension of the Hamilton approach to full QCD at finite temperature
and baryon density.

\end{document}